\documentstyle[12pt,epsf]{article}
\def\appendix#1{
  \addtocounter{section}{1}
  \setcounter{equation}{0}
  \renewcommand{\thesection}{\Alph{section}}
 \section*{Appendix \thesection\protect\indent \parbox[t]{11.715cm} {#1}}
  \addcontentsline{toc}{section}{Appendix \thesection\ \ \ #1}
  }

\newcommand{\newsection}{
\setcounter{equation}{0}
\section}

\def\bea{\begin{eqnarray}}
\def\eea{\end{eqnarray}}
\def\be{\begin{equation}}
\def\ee{\end{equation}}
\newcommand{\tr}[1]{\:{\rm tr}\,#1}

\def\const{{\rm const}}
\def\e{{\,\rm e}\,}
\def\d{\partial}
\def\D{\delta}

\newcommand{\rf}[1]{(\ref{#1})}
\newcommand{\non}{\nonumber \\*}
\def\N{{\cal N}=4}
\def\a{\alpha'}
\def\al{\alpha}
\def\ep{\epsilon}
\def\t{\theta}
\def\vt{\phi}
\def\A{{\cal A}}
\def\dl{\left\langle W^\dagger(C_1)W(C_2)\right\rangle_{{\rm conn}}}

\begin{document}
\begin{titlepage}
\begin{flushright}
ITEP--TH--17/99
\end{flushright}
\vspace{1.5cm}

\begin{center}
{\LARGE Wilson Loop Correlator} 
\\[.5cm]
{\LARGE in the AdS/CFT Correspondence}\\
\vspace{1.9cm}
{\large K.~Zarembo}\footnote{Permanent address:
{\it Department of Physics and Astronomy,}
{\it University of British Columbia,}
 {\it 6224 Agricultural Road, Vancouver, B.C. Canada V6T 1Z1} 
and
{\it Institute of Theoretical and Experimental Physics,}
 {\it B. Cheremushkinskaya 25, 117259 Moscow, Russia}.\\
E-mail: {\tt zarembo@theory.physics.ubc.ca}}
\\
\vspace{24pt}
{\it The Niels Bohr Institute\\
Blegdamsvej 17\\
DK-2100 Copenhagen 0\\
Denmark}
\end{center}
\vskip 2 cm
\begin{abstract}
The AdS/CFT correspondence predicts a phase transition in
Wilson loop correlators in the strong coupling $\N$, $D=4$ SYM
theory which arises due to instability of the classical string
stretched between the loops. We study this transition in detail
by solving equations of motion for the string in the particular case of
two circular Wilson loops.
The transition is argued to be smoothened at finite `t~Hooft coupling 
by fluctuations of the string world sheet and to be promoted to a
sharp crossover. Some general comments about Wilson loop correlators 
in gauge theories are made.
\end{abstract}

\end{titlepage}

\setcounter{page}{2}

\newsection{Introduction}

According to \cite{Mal98,Rej98}, 
calculation of Wilson loop correlators in $D=4$, $\N$ 
supersymmetric $SU(N)$ Yang-Mills theory at large $N$ and large 
't~Hooft coupling amounts in evaluation of the classical string action 
in anti-de-Sitter space. The string propagates in the bulk 
of $AdS_5\times S^5$ 
with its ends attached to the Wilson loops lying on the boundary.
This prescription is a consequence of the AdS/CFT correspondence 
\cite{Mal97,Gub98,Wit98}.
Gross and Ooguri pointed out that it implies a kind of 
phase transition in the correlation function of two Wilson loops 
\cite{Gro98a}.    
The reason for the Gross-Ooguri phase transition is that the string action, 
the area of a minimal surface bounded by the loops, generically has
two competing saddle points. The minimal surface can have a topology of
annulus or can consist of two disconnected pieces spanning individual loops.
The annulus evidently has smaller area when the loops are closed enough
to one another. But the area of the 
annulus increases with separation between 
the loops and eventually disconnected surface becomes energetically
more favourable.  At large distances, the string behaves classically only
in the vicinity of the loops and the connected correlator is saturated by 
perturbative exchange of lightest supergravity modes 
between disconnected pieces of the classical world sheet \cite{Gro98a,Ber98}. 
A jump from one saddle point to the other should lead to the phase
transition in the Wilson loop correlator 
considered as a function of the separation.

\begin{figure}[t]
\hspace*{5cm}
\epsfxsize=7cm
\epsfbox{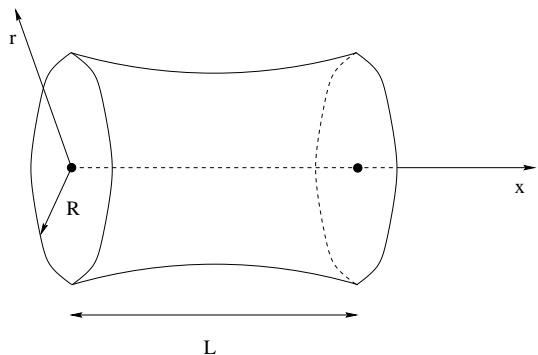}
\caption[x]{Minimal surface spanned by two concentric discs.}
\label{cat}
\end{figure}

Two Wilson loop correlators in the AdS/CFT correspondence were studied at 
large distances when they are saturated by supergraviton exchange \cite{Ber98}.
To our knowledge, 
the connected solution for the minimal surface has not yet
been considered.  
The arguments of \cite{Gro98a} relied upon the pattern of topology
change for the surface stretched between two concentric circles in the 
flat space, fig.~\ref{cat}. For further references, we sketch these arguments
here.
 
When the distance $L$ between the circles is small, the minimal surface 
is 
the catenoid: $r(x)=R_m\cosh(x/R_m)$.
The area of the catenoid is given by
\be\label{am}
A=\pi R_mL+\pi R_m^2\sinh\left(\frac{L}{R_m}\right),
\ee
where $R_m$, the radius of its narrowest section,  is 
determined by boundary conditions:
\be\label{rm}
R_m\cosh\left(\frac{L}{2R_m}\right)=R.
\ee 
The left hand side of this equation, as a function of $R_m$, 
has a minimum at some $R_m=\const\cdot L$, 
so for large enough $L$ this equation ceases
to have any solutions. At the critical distance $L_*$, the connected
minimal surface becomes unstable.
Yet before its area starts to exceed the area of two discs, another
minimal surface with the same boundary.

In this paper, we address the issue of how the Gross-Ooguri phase
transition is affected by the geometry of
$AdS_5$ and by quantum corrections due to fluctuations of the string 
world sheet. The former problem is much simpler.
The solution for two circular Wilson loops  
is found in sec.~\ref{ads}.
The surface with the topology of annulus
 becomes unstable at certain value of $L$. As a consequence,
the Wilson loop correlator in $\N$ SYM undergoes the phase
transition at infinite 't~Hooft coupling.

The influence of world-sheet fluctuations is more 
difficult to estimate. It is still not known how to calculate superstring
amplitudes in $AdS_5\times S^5$. Presence of 
the background Ramond-Ramond flux compels
one to use the Green-Schwarz formalism
\cite{Met98,Kal98a,Pes98,Kal98b,Kal98c,Raj99}, the covariant 
quantisation of the superstring 
in which is notoriously complicated \cite{GSW}. 
In view of the difficulties with passing to finite $\a$
in $AdS$ geometry, which is equivalent to
finite 't~Hooft coupling on the SYM side, we pursue the above
flat space example 
to study the world-sheet fluctuations in
sec.~\ref{flat}. The string amplitude between two loops can be calculated
non-perturbatively in $\a$ in this case \cite{Coh86,Mar89}. 

\newsection{Two Loop Correlator}\label{ads}

In the supergravity (strong coupling) limit, the Wilson loop
correlator is expressed in terms of an area of the classical string 
world sheet stretched between the loops \cite{Mal98,Rej98}:
\be\label{mal1}
\dl
=\exp\left(-\frac{S}{2\pi\a}\right),
\ee
\be\label{mal2}
S=\int d\sigma d\tau\,\sqrt{\det_{ab}g_{\mu\nu}\d_a x^\mu\d_b x^\nu},
\ee
\be\label{mal3}
\frac{\D S}{\D x^\mu}=0.
\ee
We take $C_1$ and $C_2$ to be concentric circles of radius $R$ separated
by distance $L$ (fig.~\ref{cat}). This configuration suggests the use
of the cylindric coordinates in ${\bf R}^4$. 
The $AdS_5$ metric is then
\be
ds^2=\frac{1}{z^2}\left(dz^2+dt^2+dx^2+dr^2+r^2d\varphi^2\right).
\ee
The boundary of $AdS$ is at $z=0$. We chose the units in which the
radius of $AdS$ is unity and the string tension is proportional to the 
Yang-Mills coupling: $$\a=(2g_{\rm YM}^2N)^{-1/2}.$$    

\subsection{Minimal Surface}

Using symmetries of the problem we take the following ansatz for
the minimal surface:
\be
t=0,~~\varphi=\sigma,~~r=r(\tau),~~x=x(\tau),~~z=z(\tau).
\ee
Then the equations $\D S/\D t=0$ and $\D S/\D\varphi=0$ are satisfied 
identically. The equations for $r$, $x$, and $z$ follow from the
action
\be
S=2\pi\int d\tau\,\frac{r}{z^2}\,\sqrt{(r')^2+(x')^2+(z')^2}.
\ee
The variation with respect to $r$ and $z$ yields:
\bea
\left(\frac{r}{z^2}\,\frac{r'}{\sqrt{(r')^2+(x')^2+(z')^2}}\right)'
-\frac{1}{z^2}\,\sqrt{(r')^2+(x')^2+(z')^2}&=&0,
\\*
\left(\frac{r}{z^2}\,\frac{z'}{\sqrt{(r')^2+(x')^2+(z')^2}}\right)'
+\frac{2r}{z^3}\,\sqrt{(r')^2+(x')^2+(z')^2}&=&0.
\eea
The equation which follows from the variation with
respect to $x$ can be integrated once to give
\be
\frac{r}{z^2}\,\frac{x'}{\sqrt{(r')^2+(x')^2+(z')^2}}=k,
\ee
where $k$ is an integration constant. If $k=0$, 
the string does not propagate along the $x$ direction and sweeps out
a surface with the disc topology. The minimal surface
in $AdS_5$ bounding the single 
circular Wilson loop was found in \cite{Gro98b,Ber98}:
\be\label{disc}
r^2+z^2=R^2.
\ee

If $k\neq 0$ (without loss of generality we assume that $k>0$), $x'$ is
always positive and we can choose $x$ to be one of the coordinates on the 
string world sheet: $\tau=x$. With this gauge choice,
the equations of motions take the form: 
\bea
r''-\frac{r}{k^2z^4}&=&0, \label{one}
\\*
z''+\frac{2r^2}{k^2z^5}&=&0, \label{two}
\\*
(z')^2+(r')^2+1-\frac{r^2}{k^2z^4}&=&0. \label{three}
\eea
Boundary conditions for these equations are
\be
r(-L/2)=r(L/2)=R,
\ee
\be
z(-L/2)=z(L/2)=0.
\ee
In principle, it is necessary to regularise the problem by shifting the
Wilson loops from the boundary of $AdS$ in order to get the solution with
a finite area. Strictly speaking, 
more appropriate boundary conditions for $z$ are 
$z(\pm L)=\ep$, but the solution itself is not singular in the limit 
$\ep\rightarrow 0$ and, since a regularisation does not influence the
critical behaviour, we shall first find unregularised solution and shall
return to the issue of regularisation later.

\begin{figure}[t]
\hspace*{5cm}
\epsfxsize=7cm
\epsfbox{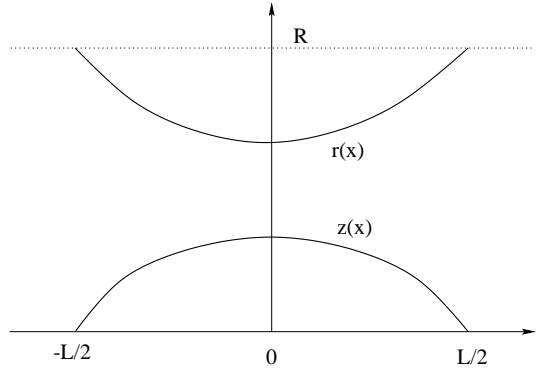}
\caption[x]{Schematic form of the solution for the minimal surface.}
\label{sol}
\end{figure}

Qualitative structure of the solution is clear from eqs.~\rf{one}, \rf{two}.
Closed string emitted by the Wilson loop at $x=-L/2$ falls into the interior 
of the $AdS$ space, then bounces back, and is absorbed by the Wilson loop
at $x=L/2$. 
The radius of the string decreases till the bouncing point and
then starts to increase (fig.~\ref{sol}), like in the flat space.

It is possible to integrate equations \rf{one}--\rf{three}. Adding first
two ones multiplied by $r$ and by $u$, respectively, to the third
equation, we get:
\be
(r^2+z^2)''+2=0,
\ee
or, taking into account boundary conditions:
\be\label{sph}
r^2+z^2+x^2=R^2+\frac{L^2}{4}\equiv a^2.
\ee
This equality suggests the substitution:
\bea
r&=&\sqrt{a^2-x^2}\cos\t, 
\non
z&=&\sqrt{a^2-x^2}\sin\t. \label{trig} 
\eea
The  equation~\rf{three} considerably simplifies in the new
variables:
\be
(a^2-x^2)(\t ')^2+\frac{x^2}{a^2-x^2}+1
-\frac{\cos^2\t}{k^2(a^2-x^2)\sin^4\t}=0.
\ee
The variables separate:
\be\label{tetash}
\t '=\pm\frac{a}{a^2-x^2}\,\sqrt{\frac{\cos^2\t}{k^2a^2\sin^4\t}-1},
\ee
where the upper sign is to be taken for $x\in [-L/2,0]$ and the lower sign
for $x\in [0,L/2]$. The solution on the interval $[-L/2,0]$  
is determined by equation
\be\label{tx}
ka\int_0^\t \,\frac{d\vt\,\sin^2\vt}{\sqrt{\cos^2\vt-k^2a^2\sin^4\vt}} 
=\frac12\,\ln\left(\frac{a+\frac{L}{2}}{a-\frac{L}{2}}\right)
-\frac12\,\ln\left(\frac{a-x}{a+x}\right)
\ee
and should be continued to positive $x$ by the symmetry 
$\t(x)=\t(-x)$.

The parameter $k$ is fixed by requirement of
continuity of the solution at $x=0$.  
Since $r$ and $z$ are even functions of $x$, so is $\t(x)$. Hence, the 
derivative $\t'(0)$ vanishes. Equation \rf{tetash} then fixes 
 the value of $\t$ at $x=0$:
\be\label{to}
\t_0\equiv \t(0)=\arccos\left(\frac{\sqrt{4k^2a^2+1}-1}{2ka}\right).
\ee
Substituting $x=0$ in eq.~\rf{tx}, we get an equation for $k$:
\be\label{fka}
F(ka)=\frac12\,\ln\left(\frac{a+\frac{L}{2}}{a-\frac{L}{2}}\right)
=\ln\left(\frac{\sqrt{R^2+\frac{L^2}{4}}+\frac{L}{2}}{R}\right),
\ee
where
\be\label{defk}
F(ka)\equiv ka\int_0^{\t_0} \,
\frac{d\vt\,\sin^2\vt}{\sqrt{\cos^2\vt-k^2a^2\sin^4\vt}}.
\ee
The right hand side is the  complete elliptic integral, since
the upper bound of integration is at the square root branching point:
$\cos^2\t_0-k^2a^2\sin^4\t_0=0$. 
The equations \rf{trig}, \rf{tx}, and \rf{fka} 
express the solution for the minimal surface in terms of elliptic 
functions.

\subsection{Critical behaviour}

The existence of the phase transition in the Wilson loop correlator can
be inferred from eq.~\rf{fka}. The function $F(\xi)$ turns to zero at 
$\xi=0$ and at $\xi=\infty$:
\bea
F(\xi)&\simeq& -\xi\ln\xi~~~(\xi\rightarrow 0),
\non\label{fxi}
F(\xi)&\simeq&\frac{\sqrt{2\pi^3}}{\Gamma^2\left(\frac14\right)}
\,\frac{1}{\sqrt{\xi}}~~~(\xi\rightarrow\infty).
\eea
Consequently, it has a maximum at some $\xi=\xi_*$. 
At the same time, the right hand side of equation \rf{fka} varies
from zero to infinity as the separation between loops increases. 
Continuity arguments show that the branch
with larger $ka$ should be chosen from the two of the 
solutions of equation
\rf{fka}. The two branches meet at $ka=\xi_*$. If $L$
exceeds the critical value
$$L_*=2R\sinh F(\xi_*),$$
the equation \rf{fka} does not have any solutions and the minimal surface
with the topology of annulus ceases to exist. Numerically, $\xi_*=0.58$
and
\be
L_*=1.04 R.
\ee   
The connected minimal surface  becomes unstable
when Wilson loops are separated by this distance. In fact, the transition
from the annulus to the disc topology is of the first
order. The critical point is determined by the equality of the free energies
in the two phases,
the areas of the disconnected and the connected
minimal surfaces. 

The area of regularised disconnected solution is \cite{Gro98b,Ber98} 
\be
S_{\rm disc}=2\left(\frac{2\pi\sqrt{R^2+\ep^2}}{\ep}-2\pi\right)
=\frac{4\pi R}{\ep}-4\pi+O(\ep).
\ee 

The area of the connected surface also requires regularisation. Shift in the
boundary conditions for $z$, $z(\pm L/2)=\ep$ instead of $z(\pm L/2)=0$,
renders the area finite. The boundary conditions for $\t(x)$
change, accordingly. From \rf{trig} we find:
\be
\t(\pm L/2)=\arctan\left(\frac{\ep}{R}\right)\simeq\frac{\ep}{R}.
\ee
Using equations of motion, we get for the regularised area:
\bea
S&=&2\pi\int_{-L/2}^{L/2}dx\,\frac{r}{z^2}\,\sqrt{1+(r')^2+(z')^2}
=\frac{2\pi}{k}\int_{-L/2}^{L/2}dx\,\frac{r^2}{z^4}
=\frac{4\pi}{k}\int_{-L/2}^{0}\frac{dx}{a^2-x^2}\,\frac{\cos^2\t}{\sin^4\t}
\non
&=&4\pi\int_{\ep/R}^{\t_0}\frac{d\t\,\cot^2\t}{\sqrt{\cos^2\t-k^2a^2
\sin^4\t}}.
\eea
This integral can be simplified by the change of variables
$$\tan\t=\left(\frac{\sqrt{4k^2a^2+1}-1}{2}\right)^{-1/2}\sin\psi.$$
After some algebra, we obtain:
\be\label{ar}
S=\frac{4\pi R}{\ep}-4\pi\,\frac{\al}{\sqrt{\al-1}}
\int_0^{\pi/2}\frac{d\psi}{1+\al\sin^2\psi+\sqrt{1+\al\sin^2\psi}},
\ee
where
\be\label{al}
\al=\frac{1+2k^2a^2+\sqrt{1+4k^2a^2}}{2k^2a^2}.
\ee
The equations \rf{mal1}, \rf{fka}, \rf{defk}, \rf{ar}, and \rf{al}
determine the correlator of two Wilson loops in the strong-coupling
$\N$ SYM theory.

The divergent part of the area is the same for the connected and 
the disconnected 
solutions. Its origin can be attributed to perimeter divergency of a
Wilson loop due to self-energy contributions \cite{Gro98b,Ber98}. 
The renormalised area
is always negative. Its short-distance asymptotics can be
found with the help of eq.~\rf{fxi}:
\be\label{asym}
S\simeq\frac{4\pi R}{\ep}-4\pi\,\frac{\sqrt{2\pi^3}}
{\Gamma^2\left(\frac14\right)}
\,ka\simeq
\frac{4\pi R}{\ep}-\frac{16\pi^4}{\Gamma^4\left(\frac14\right)}\,
\frac{R}{L}.
\ee
The area grows with separation between loops and at the point of 
the phase transition meets the area of the disconnected surface.
Solving equations for the critical point numerically, we find:
\be
L_c=0.91 R.
\ee 

In spite of evident differences between the classical string propagation 
in $AdS_5$ and in the flat space, the qualitative
pattern of the Gross-Ooguri 
phase transition appears to be not much influenced by geometry of the
target space. In both cases, the transition is of the
first order and takes place at $L\sim R$. We expect that an influence 
of the string fluctuations on the phase
transition is also universal, at least in some respects. This is our
motivation to study the free superstring propagator
between two loops in the next section.   

\newsection{Two Loop Amplitude in the Free String Theory}
\label{flat}

The equations \rf{mal1}--\rf{mal3} are expected to be valid at infinite
't~Hooft coupling and to be replaced by a full sum over random surfaces
in $AdS_5\times S^5$ with boundary conditions set by Wilson loops, 
when the coupling is finite. Presently, fluctuations of the string
world sheet in $AdS$ geometry 
can be accounted for only to the first order in 
$\a$ \cite{Gre99,For99}. The superstring 
amplitude in the flat space is much easier to calculate. It
is known exactly and provides a possibility to trace 
qualitative changes brought in by non-perturbative $\a$ corrections.     

The amplitude for the free type IIB superstring to propagate between two
concentric circular loops is a particular case of the boundary state
amplitude calculated for the bosonic string in \cite{Coh86}
and for NSR string in \cite{Mar89}. 
After GSO projection, the amplitude takes the form:
\bea\label{ampl}
\A&=&\const\,\int_0^\infty\frac{ds}{s^5}\,
\frac{\Theta_2^4(0|is)+\Theta_3^4(0|is)-\Theta_4^4(0|is)}{2\eta^{12}(is)}\,
\exp\left(-\frac{S(s)}{2\pi\a}\right)
\non
&=&\const\,\int_0^\infty\frac{ds}{s^5}\,
\frac{\Theta_2^4(0|is)}{\eta^{12}(is)}\,
\exp\left(-\frac{S(s)}{2\pi\a}\right),
\eea
where
\be\label{strac}
S(s)=\frac{L^2}{s}+2\pi R^2 \tanh\left(\frac{\pi s}{2}\right).
\ee

Suppose that $L^2\gg\a$ and $R^2\gg\a$, then $S$ is large and
the proper time integral \rf{ampl} is saturated by a saddle point,
if the latter exists. The saddle point of the action \rf{strac} is
determined by equation
\be\label{smsp}
\frac{\cosh\left(\frac{\pi s_m}{2}\right)}{\pi s_m}=\frac{R}{L}.
\ee
By the substitution $\pi s_m=L/ R_m$ we recover the boundary condition
\rf{rm} for the catenoid from this equation. The 
area of the catenoid \rf{am} is also reproduced:
\be
S(s_m)=\frac{L^2}{s_m}+\frac{L^2}{\pi s_m^2}\,\sinh \pi s_m.
\ee
The saddle point disappears 
at sufficiently large $L$, because eq.~\rf{smsp} has
no solutions for $L>L_*$. Beyond the critical point,
the main contribution to the integral comes
from $s\rightarrow\infty$. The action at $s=\infty$ is $2\pi R^2$, the area
of two discs. At large distances, the amplitude is saturated by an exchange
of massless string modes:
\be
\A\simeq \const\,\frac{96}{L^8}\,\e^{-R^2/\a}.
\ee 

\begin{figure}[t]
\hspace*{3cm}
\epsfxsize=10cm
\epsfbox{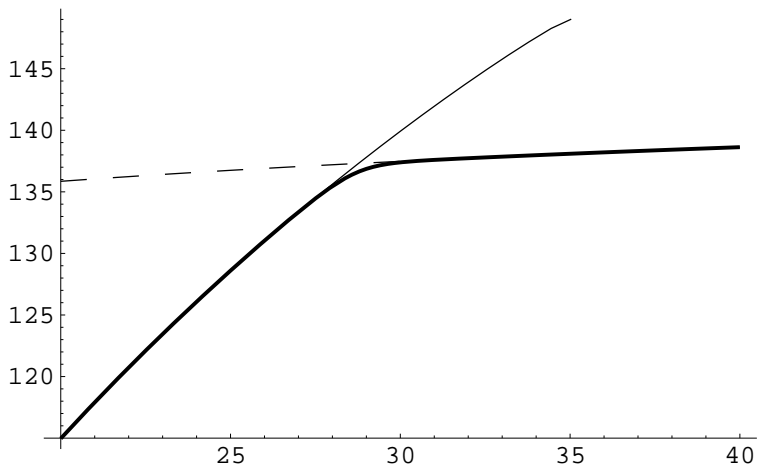}
\caption[x]{The superstring amplitude between two circular loops versus $L^2$
at $R^2=20$ (the units are such that $2\pi\a=1$).
The thick solid line is $-\ln\A$. Two other curves represent approximations 
for the amplitude valid in different phases. The thin solid line is 
the area of the catenoid with $O(\a^0)$ correction added. 
The line terminates at the point of instability, $L=L_*^2$. The dashed line
is $2\pi R^2-\ln(\mbox{supergraviton~exchange})$. The estimate
for the crossover point based on matching of the classical actions in
the two phases is $L_c\simeq 22$ for the chosen value of $R$. Account of 
the supergraviton
exchange and of the first order in the semiclassical expansion
shift the crossover point by about $20\%$. The reason for numerically
large deviation from the classical estimate is that the first order
 corrections are logarithmic in $L^2$.}
\label{amp}
\end{figure}

The semiclassical amplitude has a discontinuous
first derivative in $L$ at the point of the transition between the
two saddle points.
The discontinuity persists to any finite order in the $\a$ expansion. 
The integral \rf{ampl}, however, defines an analytic function of $L$. 
The phase transition is thus replaced by a smooth 
crossover in the exact amplitude. 
The larger $R^2$ is, the sharper this crossover will be.
The abrupt change in the amplitude is clearly seen in fig.~\ref{amp} which
displays the result of numerical integration of eq.~\rf{ampl} 
at $R^2=40\pi\a$. Note that retaining only massless string modes is 
a good approximation everywhere above the transition, since
$L^2$ is much larger than $\a$, whereas below the
transition all modes give comparable contribution. 

The conclusion from the above consideration is that
the world-sheet fluctuations smoothen the Gross-Ooguri
phase transition. However, if $\a=(2g_{\rm YM}^2N)^{-1/2}$ is sufficiently
small, the transition should show up as a sharp crossover in the Wilson loop
correlator. 

\newsection{Discussion}

Calculations carried out in the previous sections 
demonstrate that the Wilson loop two point function in $\N$ SYM undergoes
the first order phase transition as the distance between loops changes,
if the 't~Hooft coupling is infinite. If the coupling is large but finite,
tunnelling between the saddle 
points of the string action smoothens the dependence of the correlator on
the distance and the phase transition is replaced by a sharp crossover.
The transition seems to be completely washed out in the 
weak coupling, perturbative regime.

%The qualitative picture of the Gross-Ooguri phase transition depends very
%little on the details of the background metric. Therefore, the 
%existence of this transition can be regarded as a universal feature of 
%supergravity duals to strong coupling gauge theories.

In conclusion, we would like to comment on
Wilson loop correlation functions in asymptotically free, confining
gauge theories, which are also believed to have some kind of approximate, or 
even exact string representation. At short distances, 
$\Lambda_{QCD}^{-1}\ll L\ll R$, the Wilson loop correlator should be 
dominated by an open string stretched between the loops. 
This representation suggests that the large-radius limit of
the correlator should determine an interaction potential between probe 
static charges: 
\be
V(L)=-\lim_{R\rightarrow\infty}\frac{\ln\dl}{2\pi R}.
\ee
The 
comparison of eq.~\rf{asym} with the potential found in \cite{Mal98}
shows that
this relation is satisfied by the two loop correlator calculated using
the AdS/CFT correspondence. 

At large distances between loops, a more appropriate picture is
that of the closed string exchange. 
It is not quite clear whether the open and the
closed string regimes are
separated by the Gross-Ooguri 
crossover. General arguments seem to indicate that 
this is the case. If geometric characteristics of the 
loops are large in the units of the string tension, a confining
string has a large action and hopefully can be described
within the semiclassical approximation.  The instability of the classical
string world sheet will then lead to the phase transition 
in the semiclassical amplitude. This kind of behaviour certainly
holds for the free string.

However, the free string amplitude is rather loose substitute for Wilson loop
correlation functions in a gauge theory, basically because of the lack of
zig zag invariance \cite{Pol97,Pol98}. 
One of the consequences of the violation of zig zag
invariance is the Hagedorn transition in the free string case:
The amplitude \rf{ampl} diverges at
$L^2\leq L_H^2=2\pi\a$ due to exponential growth of the closed string
density of states.
Such kind of divergency of correlation functions at short distances
is expected in quantum gravity \cite{Aha99}, but not in gauge theories.
In fact,  intermediate states 
in the spectral representation
for the Wilson loop correlator,
\be\label{sr}
\dl=\int_0^\infty dE\,\rho_C(E)\e^{-EL},
\ee
\be
\rho_C(E)=\sum_{n\neq 0}\D(E-E_n)\left|\langle 0|W(C)| n\rangle\right|^2,
\ee
which generically are glueballs,
are expected to have exponentially growing spectrum. But zig zag symmetry 
suppresses coupling of Wilson loops to glueballs of very high spin, 
which can be seen
expanding the Wilson loop in local operators \cite{Shi80}. 
Spin $2n$ operators come with
coefficients of order $1/n!$. For example, the operator with 
a minimal number of derivatives is 
$$\frac{1}{n!}\,\tr\left(\oint dx_\mu x_\nu\,F_{\mu\nu}(x_0)\right)^{n}.$$
Hence, Wilson loop effectively couples only to states with spin $J<J_0$,
where $J_0$ is proportional to the area of the loop. As a result, the
spectral density $\rho_C(E)$ should have power-law asymptotics. 
This property is essentially a consequence of the 
zig zag invariance, since
the latter requires the Wilson loop of zero area to be
the unit operator. This is not true
for the boundary states in \rf{ampl}, which couple to all string modes
even at $R=0$ \cite{Coh86,Mar89}. 

The modular transformation converts the Hagedorn divergency of the
amplitude
to the tachyon singularity in the open string channel \cite{Ole86}.
Again, the tachyon instability is not expected to arise in gauge theories.

\subsection*{Acknowledgments}

I am grateful to Y.~Makeenko, P.~Olesen and G.~Semenoff for discussions
and to J.~Abmj{\o}rn for hospitality at The Niels Bohr Institute.
This work was supported by NATO Science Fellowship and, in part, by
 INTAS grant 96-0524,
 RFFI grant 97-02-17927
 and grant 96-15-96455 for the promotion of scientific schools.

%\setcounter{section}{0}
%\appendix{}


\begin{thebibliography}{99}
\addtolength{\itemsep}{-6pt}

\bibitem{Mal98}
J.~Maldacena,
``Wilson loops in large $N$ field theories,"
Phys. Rev. Lett. {\bf 80}, 4859 (1998)
hep-th/9803002.
%%CITATION = PRLTA,80,4859;%%

\bibitem{Rej98}
S.~Rey and J.~Yee,
``Macroscopic strings as heavy quarks in large $N$ gauge theory and anti-de
                  Sitter supergravity,"
hep-th/9803001.
%%CITATION = HEP-TH 9803001;%%

\bibitem{Mal97}
J.~Maldacena,
``The Large N limit of superconformal field theories and supergravity,"
Adv. Theor. Math. Phys. {\bf 2}, 231 (1998)
hep-th/9711200.
%%CITATION = 00203,2,231;%%

\bibitem{Gub98}
S.S.~Gubser, I.R.~Klebanov and A.M.~Polyakov,
``Gauge theory correlators from noncritical string theory,"
Phys. Lett. {\bf B428}, 105 (1998)
hep-th/9802109.
%%CITATION = PHLTA,B428,105;%%

\bibitem{Wit98}
E.~Witten,
``Anti-de Sitter space and holography,"
Adv. Theor. Math. Phys. {\bf 2}, 253 (1998)
hep-th/9802150.
%%CITATION = 00203,2,253;%%

\bibitem{Gro98a}
D.J.~Gross and H.~Ooguri,
``Aspects of large $N$ gauge theory dynamics as seen by string theory,"
Phys. Rev. {\bf D58}, 106002 (1998)
hep-th/9805129.
%%CITATION = PHRVA,D58,106002;%%

\bibitem{Ber98}
D.~Berenstein, R.~Corrado, W.~Fischler and J.~Maldacena,
``The Operator product expansion for Wilson loops and surfaces in the large $N$
                  limit,"
hep-th/9809188.
%%CITATION = HEP-TH 9809188;%%

\bibitem{Met98}
R.R.~Metsaev and A.A.~Tseytlin,
``Type IIB superstring action in $AdS_5\times S^5$  background,"
Nucl. Phys. {\bf B533}, 109 (1998)
hep-th/9805028.
%%CITATION = NUPHA,B533,109;%%

\bibitem{Kal98a}
R.~Kallosh, J.~Rahmfeld and A.~Rajaraman,
``Near horizon superspace,"
JHEP {\bf 09}, 002 (1998)
hep-th/9805217.
%%CITATION = JHEPA,9809,002;%%

\bibitem{Pes98}
I.~Pesando,
``A kappa fixed fixed type IIB superstring action on $AdS_5\times S^5$,"
JHEP {\bf 11}, 002 (1998)
hep-th/9808020.
%%CITATION = JHEPA,9811,002;%%

\bibitem{Kal98c}
R.~Kallosh and J.~Rahmfeld,
``The GS string action on $AdS_5\times S^5$,"
Phys. Lett. {\bf B443}, 143 (1998)
hep-th/9808038.
%%CITATION = PHLTA,B443,143;%%

\bibitem{Kal98b}
R.~Kallosh and A.A.~Tseytlin,
``Simplifying superstring action on $AdS_5\times S^5$,"
JHEP {\bf 10}, 016 (1998)
hep-th/9808088.
%%CITATION = JHEPA,9810,016;%%

\bibitem{Raj99}
A.~Rajaraman and M.~Rozali,
``On the quantization of the GS string on $AdS_5\times S^5$,"
hep-th/9902046.
%%CITATION = HEP-TH 9902046;%%

\bibitem{GSW}
M.B.~Green, J.H.~Schwarz and E.~Witten, {\sl
Superstring theory, vol.~1}, (Cambridge University Press, 1987).

\bibitem{Coh86}
A.~Cohen, G.~Moore, P.~Nelson and J.~Polchinski,
``An Off-Shell Propagator For String Theory,"
Nucl. Phys. {\bf B267}, 143 (1986).
%%CITATION = NUPHA,B267,143;%%

\bibitem{Mar89}
A.V.~Marshakov,
``The Path Integral Representation Of Fermionic String Propagator,"
Nucl. Phys. {\bf B312}, 178 (1989).
%%CITATION = NUPHA,B312,178;%%

\bibitem{Gro98b}
D. Gross, ``Aspects of Large $N$ Gauge Theory Dynamics
                        as Seen by String Theory,'' in: {\it Strings '98},
eds. S.B. Giddings, H. Ooguri, A.W. Peet and J.H. Schwarz,
{\tt http://www.itp.ucsb.edu/\~{}strings98}.
 
\bibitem{Gre99}
J.~Greensite and P.~Olesen,
``World sheet fluctuations and the heavy quark potential in the AdS / CFT
                  approach,"
JHEP {\bf 04}, 001 (1999)
hep-th/9901057.
%%CITATION = JHEPA,9904,001;%%

\bibitem{For99}
S.~Forste, D.~Ghoshal and S.~Theisen,
``Stringy corrections to the Wilson loop in N=4 superYang-Mills theory,"
hep-th/9903042.
%%CITATION = HEP-TH 9903042;%%

\bibitem{Pol97}
A.M.~Polyakov,
``String theory and quark confinement,"
Nucl. Phys. Proc. Suppl. {\bf 68}, 1 (1998)
hep-th/9711002.
%%CITATION = NUPHZ,68,1;%%

\bibitem{Pol98}
A.M.~Polyakov,
``The Wall of the cave,"
Int. J. Mod. Phys. {\bf A14}, 645 (1999)
hep-th/9809057.
%%CITATION = IMPAE,A14,645;%%

\bibitem{Aha99}
O.~Aharony and T.~Banks,
``Note on the quantum mechanics of M theory,"
JHEP {\bf 03}, 016 (1999)
hep-th/9812237.
%%CITATION = JHEPA,9903,016;%%

\bibitem{Shi80}
M.A.~Shifman,
``Wilson Loop In Vacuum Fields,"
Nucl. Phys. {\bf B173}, 13 (1980).
%%CITATION = NUPHA,B173,13;%%

\bibitem{Ole86}
P.~Olesen,
``On The Exponentially Increasing Level Density In String Models And The
                  Tachyon Singularity,"
Nucl. Phys. {\bf B267}, 539 (1986).
%%CITATION = NUPHA,B267,539;%%

\end{thebibliography}
\end{document}